\documentclass[12pt]{article}
\usepackage{amssymb,epsfig}

\renewcommand{\thefootnote}{\fnsymbol{footnote}}

\begin{document}

\title{
\begin{flushright}
\ \\*[-80pt] 
\begin{minipage}{0.3\linewidth}
\normalsize
arXiv:0804.3229 \\
YITP-08-28 \\
TU-816 \\
KUNS-2136 \\
IFT-UAM/CSIC-08-22\\*[50pt]
\end{minipage}
\end{flushright}
{\Large \bf 
Non-perturbative moduli 
superpotential \\ with positive exponents
\\*[20pt]}}

\author{
\centerline{
Hiroyuki~Abe$^{1,}$\footnote{
E-mail address: abe@yukawa.kyoto-u.ac.jp}, \ 
Tetsutaro~Higaki$^{2,}$\footnote{
E-mail address: tetsu@tuhep.phys.tohoku.ac.jp}, \ 
Tatsuo~Kobayashi$^{3,}$\footnote{
E-mail address: kobayash@gauge.scphys.kyoto-u.ac.jp} \ and \ 
Osamu~Seto$^{4,}$\footnote{E-mail address: osamu.seto@uam.es}} 
\\*[20pt]
\centerline{
\begin{minipage}{\linewidth}
\begin{center}
$^1${\it \normalsize 
Yukawa Institute for Theoretical Physics, Kyoto University, 
Kyoto 606-8502, Japan} \\
$^2${\it \normalsize 
Department of Physics, Tohoku University, 
Sendai 980-8578, Japan} \\
$^3${\it \normalsize 
Department of Physics, Kyoto University, 
Kyoto 606-8502, Japan} \\
$^4${\it \normalsize 
Instituto de F\'isica Te\'orica UAM/CSIC,  \\
Universidad Aut\'onoma de Madrid, Cantoblanco, Madrid 28049, Spain} 
\end{center}
\end{minipage}}
\\*[50pt]}

\date{
\centerline{\small \bf Abstract}
\begin{minipage}{0.9\linewidth}
\medskip 
\medskip 
\small
We study non-perturbative moduli superpotentials with 
positive exponents, i.e. the form like $Ae^{aT}$ with 
a positive constant $a$ and the modulus $T$.
These effects can be generated, e.g., by D-branes which 
have negative RR charge of lower dimensional D-brane.
The scalar potentials including such terms have 
a quite rich structure.
There 
are several local minima with different potential energies and 
a high barrier, whose 
height is of ${\cal O}(M_p^4)$.
We discuss their implications from the viewpoints of 
cosmology and particle phenomenology, e.g. 
the realization of inflation models, avoiding 
the overshooting problem.
This type of potentials would be useful to 
realize the inflation and low-energy supersymmetry breaking.
\end{minipage}
}

\begin{titlepage}
\maketitle
\thispagestyle{empty}
\end{titlepage}


\renewcommand{\thefootnote}{\arabic{footnote}}
\setcounter{footnote}{0}

\section{Introduction}

Moduli fields play an important role in 
string phenomenology and cosmology.
Several couplings in 4D low-energy effective field theory 
are given as functions of vacuum expectation values (VEVs) of 
moduli.
Thus, we need a stabilization mechanism of moduli.
However, some of moduli fields $T$ have a flat potential 
perturbatively, while others may be stabilized 
by non-trivial background such as 
flux compactification \cite{Giddings:2001yu,Kachru:2003aw}.
Non-perturbative effects are assumed to stabilize such moduli.
The form of non-perturbative terms in superpotential 
behaves like $e^{-aT}$, where $T$ is a modulus field and $a$ is 
a positive constant, 
and such terms would be induced by gaugino condensation 
and/or stringy non-perturbative effects.
Such a potential may generate a hierarchically small 
energy scale compared with the Planck scale 
$M_p=2.4 \times 10^{18}~{\rm GeV}$.
That may be relevant to supersymmetry breaking 
and/or cosmological aspects, e.g. inflation 
models \cite{Kallosh:2007ig}.

The bumps generated by the above terms $e^{-aT}$ are 
not high in several models of moduli stabilization, 
in particular, in models leading to low-energy supersymmetry 
breaking.
That may lead to problems.
For example, a simple model has a local minimum leading to 
a finite VEV of $T$ and the runaway vacuum, which is 
the minimum corresponding to  $T \rightarrow \infty$, 
and there is a low bump between them.
In such a model, we need fine-tuning of initial conditions 
in order not to overshoot the minimum with a finite VEV of 
modulus \cite{Brustein:1992nk}.
Also, such a low bump may have a problem from 
the viewpoint of inflation.
Suppose that the inflaton $Z$ is different from $T$.
We need the positive vacuum energy deriving 
inflation and it would be higher than the above bump.
Then, the modulus would run away to infinity 
during the inflation.
Similarly, finite temperature effects may also destabilize 
the modulus  \cite{Buchmuller:2004xr}.
Furthermore, if the modulus $T$ is the inflaton, 
it seems difficult to realize the inflation and 
low-energy supersymmetry breaking in a simple model 
\cite{Kallosh:2004yh,Kallosh:2006dv}.

In Ref.~\cite{Abe:2005rx}, it was pointed out that 
non-perturbative terms  with positive exponents like 
$e^{aT}$, where $a >0$, can be induced in 
string-derived effective supergravity theory.
Suppose that the gauge kinetic function $f$ is 
written as $f=mS-wT$ with $m,w > 0$ and 
$S$ is stabilized with a heavy mass of ${\cal O}(M_p)$ 
by flux compactification.
Then, the gaugino condensation of such a sector 
induces a non-perturbative term like $e^{aT}$ 
for the light modulus $T$.
Some aspects of such a term, e.g. moduli stabilization 
and supersymmetry breaking, have been 
in Ref.~\cite{Abe:2005rx}.\footnote{See also 
Refs.~\cite{Choi:2006bh,Choi:2006xb}. 
In particular, Ref.~\cite{Choi:2006xb} has studied 
realization of the model solving the fine-tuning 
problem in the minimal supersymmetric standard model \cite{Choi:2005hd}.}
Here we study more about non-perturbative terms 
with positive exponents like  $e^{aT}$ ($a>0$).
They could generate high barriers in the scalar potential, whose 
heights would be of ${\cal O}(M_p^4)$.
We study implications of such terms from the viewpoints of 
cosmology and particle phenomenology, in particular, 
the overshooting problem, realization of inflation models and 
low-energy supersymmetry breaking.

This paper is organized as follows.
In section 2, we explain how non-perturbative terms 
with positive exponents like  $e^{aT}$ can be generated 
in the superpotential, and study the form of scalar 
potential.
In section 3, we study their implications on 
cosmology and particle phenomenology.
In section 4, we apply to the racetrack inflation.
Section 5 is devoted to conclusion and discussion.

\section{Non-perturbative moduli superpotential 
with positive exponents}

For concreteness, we consider a supergravity model, 
which could be derived from  type IIB superstring theory 
as its low-energy effective theory, although our supergravity 
model might be derived from other types of superstring theories 
such as type IIA superstring theory and heterotic 
string/M theory.\footnote{Definitions of modulus in type IIA 
and IIB superstring theories and heterotic string/M theory 
are different from each other.}
In well-known Calabi-Yau models, there are three types of closed string moduli fields, 
the dilaton $S$, the K\"ahler (volume) moduli and complex structure (shape)
moduli $U_\alpha$.
For simplicity, we consider the model with a single 
K\"ahler modulus $T$, but extensions to models with 
several K\"ahler moduli are straightforward.
Following Ref.~\cite{Kachru:2003aw}, we assume that 
the dilaton $S$ and complex structure moduli $U_\alpha$
are stabilized by flux-induced superpotential 
$W_{\rm flux}(S,U_\alpha)$ \cite{Gukov:1999ya}.
That implies that those moduli fields have heavy masses 
of ${\cal O}(M_p)$.
Here and hereafter we use the unit that $M_p=1$.

In order to stabilize the remaining light modulus $T$, 
one often assumes the gaugino condensation in the hidden 
sector with the gauge kinetic function $f_a$,
which induces the following term in the superpotential,
\begin{equation}
W_{\rm np} = A e^{-\frac{2 \pi}{N_a}f_a},
\end{equation} 
where $A ={\cal O}(M_p^3)$ and we have assumed that 
the hidden sector is described by a ${\cal N}=1$ pure super Yang-Mills (SYM) theory of $SU(N_a)$.
Note that we use the normalization of the holomorphic gauge 
kinetic function $f_a$ such that the gauge coupling $g_a$  at cut off scale 
is obtained as 
\begin{equation}
\frac{4\pi}{g_a^2}= Re(f_a) ,
\end{equation}
except non-holomorphic terms from $\sigma$-model anomalies.
Thus, in the simple model with the holomorphic gauge kinetic function at tree level
$f_a=T$ like the gauge sector on D7-brane {\footnote{Hereafter we use a symbol of $f_a$ as a holomorphic gauge coupling at string tree level.}}, 
the gaugino 
condensation induces{\footnote{From this equation, 
a factor of $A$ can include VEVs of complex structure moduli $U_{\alpha}$, 
whose effects come from threshold corrections to gauge coupling by heavy mode in ${\cal N}=2$ SUSY sector of open string.
For the (non-)perturbative corrections to gauge coupling, see Ref. \cite{Blumenhagen:2006ci}.}} 
\begin{equation}
W_{\rm np} = A e^{-\frac{2 \pi}{N_a}T}.
\label{eq:np-W-1}
\end{equation}

However, the gauge kinetic function is written 
as a linear combination of two or more moduli 
in several string theories like heterotic string/M 
theory \cite{Choi:1985bz} and type II string theories with 
magnetized D-branes and/or intersecting 
D-branes \cite{Blumenhagen:2006ci, Cremades:2002te,Lust:2004cx, Bertolini:2005qh}.
For example, a stack of magnetized D7-branes has the following 
gauge kinetic functions,
\begin{equation}
f_a = m_a S + w_a T,
\label{eq:gauge-f-2}
\end{equation}
where $m_a$ and $w_a$ correspond to RR charges of D3-brane and one of D7-brane respectively, which 
are carried by the single magnetized D7-brane. On the magnetized D7-brane, they 
are positive rational numbers determined by magnetic fluxes and winding numbers
{\footnote{The form of gauge couplings in this paper can be found in compactifications of toroidal orbifold at least. 
However, in generic Calabi-Yau compactifications this changes due to the geometric curvature terms. 
For example, ${\cal O}({\cal R}^2)$ terms exist in $m_a$ of
a gauge coupling on the (magnetized) D7-brane \cite{Haack:2006cy}.
Then we may have negative $m_a$ of it as magnetized D9-branes which have 
negative D3-brane charges \cite{Cascales:2003zp}. 
At any rate, we will mention the case which is independent of 
geometric curvature terms.}}. 
Similar gauge kinetic functions can be derived from 
other string theories including heterotic string/M 
theory and type IIA superstring theory.
When the gaugino condensation happens in the ${\cal N}=1$ pure SYM of $SU(N_a)$ with 
this gauge kinetic function (\ref{eq:gauge-f-2}), the following term in the superpotential 
is induced,
\begin{equation}
W_{\rm np} = A e^{-\frac{2 \pi}{N_a}(m_a S + w_a T)}.
\label{eq:np-W-2}
\end{equation} 
We assume that the dilaton $S$ is already stabilized 
with the mass of ${\cal O}(M_p)$ by the flux 
compactification.
Thus, here we replace $S$ by its VEV $S_0$.\footnote{
Such replacement by $S_0$ is valid in the case that 
the dilaton mass is much larger than the mass of 
K\"ahler modulus $T$ \cite{Choi:2004sx,deAlwis:2005tf,Abe:2006xi}.
(See e.g. Ref.~\cite{Abe:2005pi} for the model that 
both $S$ and $T$ are light moduli.)
This condition is satisfied with the cases that we study.}
Then, the superpotential reduces to the form 
$A'e^{-aT}$ with $A' = A e^{-\frac{2 \pi}{N_a}m_a S_0}$ 
and $a = \frac{2 \pi}{N_a}w_a$, 
and its form is almost the same as the 
superpotential (\ref{eq:np-W-1}).
However, the coefficient $A'$ can be hierarchically suppressed 
compared with $M_p^3$, because of the factor 
$e^{-\frac{2 \pi}{N_a}m_a S_0}$.

On the other hand, the following form of gauge 
kinetic function 
\begin{equation}
f_a = m_a S - w_a T,
\label{eq:gauge-f-3}
\end{equation}
with positive rational numbers $m_a$ and $w_a$ 
can also be derived from supersymmetric
magnetized D9-brane
which carries negative RR charge of 
D7-brane \cite{Cascales:2003zp}
{\footnote{This means that 
we have (almost) vanishing Fayet-Iliopoulos D-term on the magnetized D9-brane.
For example, in $T^6/(Z_2 \times Z_2) $ orientifold ($h_{1,1}^{(+) ~ untwist}=3$), 
the condition is proportial to the equation, e.g.,
$D \propto \frac{1}{mRe(S)} + \frac{1}{w_i Re(T_i)}+ \frac{1}{w_jRe(T_j)} -\frac{1}{w_kRe(T_k)} 
\simeq 0$ 
with $\forall~w_i ,~m>0$, while a holomorphic gauge coupling on the magnetized D-brane is given by 
$f= mS +w^i T_i+w^j T_j -w^k T_k$.
Here $i,j,k=1,2,3$, $i\neq j \neq k \neq i$ and we omitted the contributions of matter fields.
}}, 
as well as heterotic string/M 
theory and type IIA superstring theory, 
when $m_a Re(S) - w_a Re(T) >0$.\footnote{In D-brane systems
a gauge coupling at tree level on a D-brane is given by a VEV of linear combination of moduli, 
which means effective volume wrapped by the D-brane.
Especially, in type IIB O3/O7 system tree level gauge coupling would be reliable as long as 
$Re(T)/Re(S) = R^4 > 1$, where $R$ is a radius of compactification 
normalized by string length $\alpha '^{1/2}$ in string frame. In heterotic case, a (linear) 
combination of moduli appears at 1-loop (or next $\kappa_{11}^{2/3}$) order. }
As the magnetized D7-brane, $m_a$ and $-w_a$ correspond to RR charge of D3-brane and one of D7-brane respectively, 
which are carried by the single magnetized D9-brane.
The gaugino condensation in the hidden sector of ${\cal N}=1$ pure SYM of $SU(N_a)$ with
this gauge coupling induces the following 
non-perturbative term in the superpotential
{\footnote{For gauge couplings
$f^{k}_{a}= mS +w^i_a T_i + w^j_aT_j - w^k_a T_k$, where $a$ represents a label for stacks of 
the magnetized D9-branes, $i,j,k=1,2,3 ~{\rm and}~i\neq j \neq k \neq i$,
it is sufficient for us to have a superpotential, e.g., $W=  A_{1} 
e^{-a_{1}f^{1}_{1}} + A_{2} e^{-a_{2}f^{2}_{2}}+A_{3} e^{-a_{3}f^{3}_{3}}$, 
which prevents each modulus from running away to infinity, though 
in $T^6/(Z_2 \times Z_2)$ orientifold,
studies for the stabilization of open string moduli are important \cite{Blumenhagen:2005tn}.
}}, 
\begin{equation}
W_{\rm np} = A e^{-\frac{2 \pi}{N_a}(m_a S_0 - w_a T)}.
\label{eq:np-W-3}
\end{equation}
Here, we have assumed that the dilaton is stabilized 
with a heavy mass of ${\cal O}(M_p)$ and replaced 
$S$ by its VEV $S_0$.
The superpotential (\ref{eq:np-W-3}) corresponds to the form $A'e^{aT}$ with 
the positive exponent $a= \frac{2\pi w_a}{N_a}$, and that leads to 
the moduli potential quite different from 
one only with negative exponents, i.e. $A'e^{-aT}$.
The region around $ w_a T = m_a S_0$
corresponds to the strong gauge coupling 
region, where it would not be reliable to consider 
only the term $Ae^{-\frac{2\pi}{N_a}f_a}$ in the superpotential. 
The strong gauge coupling region 
would correspond to $4\pi /g^2 = Re(f_a) \lesssim 1$, i.e., 
\begin{equation}
m_a Re(S_0) - w_a Re(T) \lesssim 1.
\label{eq:strong}
\end{equation}
Outside of this region (\ref{eq:strong}) the above superpotential
(\ref{eq:np-W-3}) is reliable.

The F-term scalar potential $V_F$ is written as 
\begin{equation}
V_F = e^K[|D_TW|^2 K^{T\bar{T}} - 3|W|^2],
\end{equation}
with $D_T W \equiv K_T W + W_T$ and 
\begin{equation}
K = -3\ln (T + \bar T),
\end{equation}
where $K_T$ and $W_T$ denote first derivatives of $K$ and
$W$ by $T$, respectively.
The gravitino mass $m_{3/2}$ is obtained as 
$m_{3/2}^2 = e^K|W|^2$.
The scalar potential $V_F$ including the above superpotential
(\ref{eq:np-W-3})
has a quite high barrier around 
$w_a Re(T) \approx m_a Re(S_0)$,\footnote{Then we will need a condition that $m_a > w_a$ 
from the condition that $Re(T)/Re(S_0) >1$. For a detail, see Ref.\cite{Abe:2005rx}.} and 
its reliable height is at least $V_F \sim |A|^2e^{-4\pi/N_a}$, 
where we have estimated at $Re(f_a) \sim 1$.
Thus, this barrier height is almost of 
${\cal O}(M_p^4)$ when $4 \pi /N_a \sim 1$.
That has significant implications in 
cosmology and particle phenomenology.
We shall study them in the next sections.

\section{Potential forms and their Implications}

Here, let us study implications of the following 
total superpotential,
\begin{equation}
W_{\rm tot} = W_0 + \sum_a A_a e^{-\frac{2\pi}{N_a}(m_aS_0 + w_a T)}
+ \sum_b A_b e^{-\frac{2\pi}{N_b}{(m_bS_0 - w_b T)}},
\label{eq:w-tot}
\end{equation}
where $^{\forall} A_a, A_b = {\cal O}(M_p^3)$.
In particular, the third term is important.
The first term corresponds to the VEV of the 
flux-induced superpotential $W_{\rm flux}(S,U_\alpha)$ 
and/or non-perturbative terms including only the dilaton $S$, 
i.e., $A_a e^{-\frac{2\pi}{N_a}m_aS_0}$.

We consider the corresponding F-term scalar potential $V_F$, 
which would have several local minima.
Also we add the uplifting potential, $E/(T+\bar T)^n$
following  \cite{Kachru:2003aw} and the total potential 
is obtained as 
\begin{equation}
V = V_F + \frac{E}{(T+\bar T)^n}.
\end{equation}
Such uplifting potential can be generated by 
anti D3-brane \cite{Kachru:2003aw}.\footnote{
Similar uplifting is possible by spontaneous SUSY breaking 
sectors \cite{Saltman:2004sn,Dudas:2006gr}.}
We tune the constant $E$ 
such that one of local minima has a small positive vacuum energy, 
$V \simeq 10^{-120}$.

This potential $V$ has a quite rich structure.
First of all, the potential $V$ as well as $V_F$ has 
a barrier with height of ${\cal O}(M_p^4)$ 
around $T \approx m_bS_0/ w_b $.
Such a high barrier would be useful to avoid the 
overshooting problem and destabilization 
due to inflation  driving energy and 
finite temperature effects.
Furthermore, this potential may have several   
local minima with hierarchically different 
potential energies.
That would be useful to realize both 
inflation and low-energy supersymmetry breaking.


\subsection{Superpotential with a single term}

One of the simplest models is the model with 
the following total superpotential,
\begin{equation}
W_{\rm tot} = A e^{-\frac{2\pi}{N}{(mS-wT)}}.
\label{W-1}
\end{equation}
This superpotential is R-symmetric.
As shown in Ref.\cite{Abe:2007ax}, 
the SUSY point $D_TW=0$ corresponds to a local maximum 
of the F-term scalar potential $V_F$ and it has 
a SUSY breaking local minimum at $Re(T) = 2/a$, 
where $a=2\pi w/N$.\footnote{
A similar potential was discussed for twisted moduli \cite{Higaki:2003zk}.}
This local minimum always has a negative vacuum energy $V_F <0$.
We need the uplifting term $E/(T+\bar T)^n$ to realize a de Sitter vacuum.
Then, we require $V=V_F+E/(T+\bar T)^n \simeq 0$ and 
$\partial_T V=0$.\footnote{In order to obtain the minimum, the condition that 
$\frac{1}{2}a^3t^3 -4a^2t^2+\left(13-\frac{n(n+1)}{2} \right)at +3(n(n+1)-6) >0$ must be satisfied, too.}
These provide with the following condition,
\begin{equation}
\frac13 a^2t^2+\frac13  (n-7)at +(4-2n) =0,
\label{cond-1}
\end{equation}
where $t=2Re(T)$.
For example, for $n=2$, this condition is satisfied when 
\begin{equation}
2Re(T)=\frac{5}{a}=\frac{5N}{2\pi w},
\label{T-min}
\end{equation}
and in this case the total scalar potential is given as 
\begin{equation}
V(t) = \frac{a|A'|^2}{3t^2}[e^5+e^{at}(at-6)],
\end{equation}
where $A'=Ae^{-2\pi m S_0/N}$.
For a small value of $Re(T)$, corrections to the 
K\"ahler potential would be important.
In order to realize the minimum with $Re(T)={\cal O}(4\pi)$ from 
Eq.~(\ref{T-min}), we need 
$N/w ={\cal O}(10)-{\cal O}(100)$. 
At the minimum, SUSY is broken and 
the F-term of the modulus $T$, 
\begin{equation}
F^T= - e^{K/2}K^{T \bar T}D_TW,
\end{equation}
is evaluated as 
\begin{equation}
\frac{{F^T}}{T+\bar{T}}= -\frac{2}{3}m_{3/2},
\end{equation}
where the gravitino mass $m_{3/2}$ is given as $m_{3/2} = A'e^{5/2}/(2Re(T))^{3/2}$.
For $Re(T) = {\cal O}(4 \pi)$, the F-term $F^T$ is 
sizable compared with the anomaly mediation.
Similarly,
for $n=3$, the above condition (\ref{cond-1}) is satisfied when 
\begin{equation}
2Re(T)=\frac{2+\sqrt{10}}{a}=\frac{(2+\sqrt{10})N}{2\pi w},
\end{equation}
and the total scalar potential is given as 
\begin{equation}
V(t) = \frac{a|A'|^2}{3t^3}[2(\sqrt{10}-1)e^{2+\sqrt{10}} + e^{at}at(at-6)].
\end{equation}
The size of $F^T/(T + \bar T)$ is of ${\cal O}(m_{3/2})$, 
as the case with $n=2$.

All of the above aspects are different form the 
superpotential $W_{\rm tot} = A' e^{-aT}$ 
with $a >0$, whose scalar potential 
has no local minimum.
Thus, the total superpontential (\ref{W-1}) gives 
the simplest model for modulus stabilization, 
that is, the superpotential with a single term.
Also, this is the simplest model from the viewpoint to realize 
the modulus mediation.
(See also Ref.~\cite{Abe:2005pi}.)

The real part of $T$ can be stabilized by 
this simple R-symmetric superpotential, but 
the potentials with and without 
the uplifting term do not include 
$Im(T)$.
This aspect may be important from the viewpoint 
of the QCD axion.\footnote{
The decay constant in this model would be of 
${\cal O}(M_p)$ or order of the GUT scale. 
We would need some mechanism to lead to 
the cosmologically allowed window of the decay constant.}
(See e.g. Ref \cite{Conlon:2006tq} and references therein.)


\subsection{Superpotential with two terms: \\
the KKLT type and racetrack type}

Next, we consider models of the 
total superpotentials with only two 
terms.
Among them, the KKLT type of the 
total superpotential,
\begin{equation}
W_{\rm tot} = W_0 + A e^{-\frac{2\pi}{N}(mS_0+wT)},
\label{eq:KKLT-W}
\end{equation}
with $m \geq 0,w>0$
and the racetrack type,
\begin{equation}
W_{\rm tot} = A_1 e^{-\frac{2\pi}{N_1}(m_1S_0+w_1T)} 
+ A_2 e^{-\frac{2\pi}{N_2}(m_2S_0+w_2T)},
\end{equation}
with $m_a \geq 0,w_a>0$ $(a=1,2)$ are well-known.
The corresponding F-term scalar potentials $V_F$  
have local minima determined by $D_TW=0$,
i.e.,
\begin{equation}
Re(T) \approx  -\frac{m}{w} Re(S_0) + 
\frac{N}{2\pi w}\ln (A/W_0), 
\end{equation} 
for $\frac{2\pi w}{N}Re(T) ={\cal O}(10)$ 
in the KKLT type and 
\begin{equation}
Re(T) \approx 
\frac{-m_1N_2+m_2N_1}{w_1N_2 - w_2N_1} Re(S_0)
+ \frac{1}{2\pi(w_1/N_1 - w_2 /N_2)} 
\ln \left( \frac{A_1w_1N_2}{A_2w_2N_1} \right),
\end{equation}
for $\frac{2\pi w_a}{N_a}Re(T) ={\cal O}(10)~(a=1,2)$ in the racetrack type, respectively.
In both models, $Im(T)$ is determined 
at local minima.
These vacua remain even after we add the uplifting 
potential $E/(T + \bar T)^n$ tuning $E$ such that 
these vacua have almost vanishing energies.
Both models have the runaway behavior 
at $Re(T) \rightarrow \infty$, 
and such runaway vacuum and local minima are 
separated by not high bumps.
For example, the KKLT model has a low bump, 
whose height is of ${\cal O}(|W_0|^2)={\cal }(m_{3/2}^2M_p^2)$, 
when one tunes the constant $E$ in the uplifting potential 
such that the above local minimum has almost vanishing 
vacuum energy.
{ 
Thus, the inflaton potential energy must be lower than the height of 
${\cal }(m_{3/2}^2M_p^2)$ to aviod the runaway behavior. 
Since it means $H_{\rm inf} < m_{3/2}$ with $H_{\rm inf}$ being 
the Hubble parameter during inflation~\cite{Kallosh:2004yh}, 
inflation of a very low energy scale or a very heavy gravitino mass is required. 
For such a low energy inflation, conceptually new flatness problem is re-introduced and phenomenologically the detection of tensor type perturbation is hopeless.
On the other hand, 
very heavy gravitino is not desired from the viewpoint of 
weak scale supersymmetry as a solution to the hierarchy problem.
}
Similarly, the racetrack model has a low 
bump between the local minimum and the runaway vacuum.
Such forms of potentials would have 
problems, and one of them is the overshooting problem, 
that is, we need the fine-tuning of initial conditions 
for realizing the local minimum with a finite value of $T$.
Otherwise, the value of $Re(T)$ runs away to infinity, 
$Re(T) \rightarrow \infty$.
Furthermore, it would be difficult to 
realize inflation models in the above types of 
potentials.
We will discuss this point in section 3.3.

Now, let us study other types of the total superpotentials 
with two terms,
\begin{equation}
W_{\rm tot} = W_0 + A e^{-\frac{2\pi}{N}{(mS_0-wT)}},
\label{eq:w-total-4}
\end{equation}
with $m,w>0$ and 
\begin{equation}
W_{\rm tot} = A_1 e^{-\frac{2\pi}{N_1}(m_1S_0 +w_1T)} 
+ A_2 e^{-\frac{2\pi}{N_2}{(m_2S_0 - w_2T)}},
\label{eq:w-total-5}
\end{equation}
with $m_1 \geq 0 ,m_2,w_a>0$ $(a=1,2)$.
These may look similar to the KKLT type and 
the racetrack type of superpotentials.
However, the second terms in both superpotentials 
have positive exponents for $T$ for certain regions of $Re(T)$, 
and that leads to quite different features.
The corresponding scalar potentials $V_F$ as well as 
$V$ have barriers around $wRe(T) \approx mRe(S_0)$ and 
$w_2 Re(T) \approx m_2Re(S_0)$, respectively.
They have local minima corresponding to $D_T W =0$, i.e.,
\begin{equation}
Re (T) \approx \frac{N}{2\pi w} \ln(W_0/A) +\frac{m}{w}Re(S_0),
\end{equation}
for $W_{\rm tot}$ (\ref{eq:w-total-4}), and 
\begin{equation}
Re (T) \approx \frac{m_2N_1 - m_1N_2}{w_1N_2+w_2N_1} Re(S_0) + 
\frac{1}{2 \pi (w_1/N_1 + w_2/N_2)} 
\ln \left( \frac{A_1w_1N_2}{A_2w_2N_1}\right) ,
\end{equation}
for $W_{\rm tot}$ (\ref{eq:w-total-5}).
Since obviously there is a barrier of ${\cal O}(M_p^4)$
for small $Re(T)$ because of the K\"ahler potential
$K= -3\ln(T+\bar{T})$, 
the above local minima are surrounded by 
high barriers of ${\cal O}(M_p^4)$.
Thus, the overshooting problem would be avoided, that is, 
we do not need the fine-tuning to realize
the above local minima.
This aspect would also be useful for realization of the 
inflation as we will discuss in section 3.3.

Finally, among the class of superpotentials with 
only two terms, let us consider the following,
\begin{equation}
W_{\rm tot} = A_1 e^{-\frac{2\pi}{N_1}{(m_1S_0  -w_1T)}} 
+ A_2 e^{-\frac{2\pi}{N_2}{(m_2S_0 - w_2T)}},
\label{eq:w-total-6}
\end{equation}
with $m_a,w_a>0$ $(a=1,2)$, which satisfies 
$m_1/w_1 < m_2/w_2 $, $w_1/N_1 > w_2/N_2$ and  $m_1/N_1 > m_2/N_2$.
We restrict to the region 
$0 \lesssim Re(T) \lesssim m_1 Re(S_0)/w_1$.
That looks similar to the racetrack superpotential, but 
the sign of exponents for $T$ is opposite.
In this case, there is the local minimum,
\begin{equation}
Re(T) \approx \frac{m_1N_2 - m_2N_1}{w_1N_2 - w_2N_1} Re(S_0)
+ \frac{-1}{2\pi(w_1/N_1 - w_2 /N_2)} 
\ln \left( \frac{A_1w_1N_2}{A_2w_2N_1} \right).
\end{equation}
Note that in this region of parameters the first term is positive.

In the superpotential with three or more terms, 
the corresponding scalar potential has a richer structure.
There are several local minima with different potential 
energies in addition to the runaway vacuum and/or 
there is a high barrier, whose height would be of 
${\cal O}(M_p^4)$.
That would be useful, e.g. to realize the inflation and 
low-energy supersymmetry breaking.

\subsection{Stability during inflation}

As discussed in the previous sections, section 3.1 and 3.2, 
the behavior of 
the total superpotential with the positive exponent term $Ae^{aT}$ 
($a>0$) is quite different from 
the total superpotential without such a term.
The bump of the scalar potential without 
the positive exponent term would be much lower than the Planck scale.
That would have several problems such as the overshooting 
problem and instability due to finite temperature effects and/or 
vacuum energy deriving inflation.
On the other hand, the scalar potential 
with the positive exponent term would have a high barrier, whose 
reliable height is almost of ${\cal O}(M_p^4)$.
That would be useful to avoid the above problems.
Here, we consider the inflation model, where 
some field $Z$ other than the modulus $T$ plays a role as 
the inflaton.

Suppose that we have a supergravity inflation model 
with the inflaton $Z$, when the degree of freedom of the modulus $T$ 
is frozen (by hand).
We write its potential as $V_{\rm inf}(Z,\bar Z)$, 
and its value during the inflation would be much 
higher than heights of the above bumps of the superpotential 
without the positive exponent term, e.g. 
$V_{\rm inf}(Z, \bar Z) \gg |W_0|^2$ in (\ref{eq:KKLT-W}).
However, the inflation is difficult to be realized 
when we consider the modulus $T$ as the dynamical field, which will
spoil original inflationary dynamics, e.g. the slow-roll condition.
If the F-term is dominant in this inflation model, 
the total potential would behave like 
\begin{equation}
V(T,\bar T, Z,\bar Z) \sim V_{\rm modulus} + \frac{V_{\rm inf}(Z,\bar Z)}{(T +  \bar T)^3} 
+ \cdots,
\end{equation}
during the inflation, where $V_{\rm modulus}$ denotes the scalar
potential stabilizing $T$ when $V_{\rm inf}(Z,\bar Z)$ is absent.
Note that $V_{\rm inf}(Z,\bar Z) \gg |W_0|^2$.
Then, $T$ runs away to infinity.
If the D-term potential, $V_D=\frac{g^2}{2}D^2$ 
is dominant during the 
inflation and the $T$-dependence appears only 
through the gauge coupling $g^2 \sim 1/(T + \bar T)$, 
the total potential would behave like 
\begin{equation}
V(T,\bar T, Z,\bar Z) \sim V_{\rm modulus} + \frac{V_{\rm inf}(Z,\bar Z)}{(T +  \bar T)} 
+ \cdots,
\end{equation}
during the inflation, and in this model 
$T$ runs away to infinity.
Also, in other cases, it would be difficult 
to realize the inflation model of $Z$ 
for the dynamical $T$ with a low bump potential.

However, the situation is different when the total superpotential 
includes a positive exponent term, $Ae^{-\frac{2\pi}{N}(mS_0 - wT)}$.
The scalar potential $V_{\rm modulus}$ has a high barrier 
around $Re(T)=mRe(S_0)/w$,
and its height is of ${\cal O}(M_p^4)$.
Thus, $Re(T)$ does not run away to 
infinity during the inflation, when its initial value is smaller than 
the location of high barrier, that is, 
$Re(T)$ would be stabilized at 
$e^{-\frac{2\pi}{N}[m(S_0 + \bar S_0)-w(T+\bar T)]} \sim 
V_{\rm inf}(Z,\bar Z)$ during the inflation 
and its mass would be larger than 
the inflation Hubble value by a factor 
$(2\pi w/N)Re(T_{\rm inf}) = {\cal O}(10)$.\footnote{The $Re(T_{\rm inf})$ 
means fixed value during inflation by $Z$ field and will satisfy an inequality
that $\langle Re(T) \rangle < Re(T_{\rm inf}) \lesssim m Re(S_0)/w$.}
Hence, the positive exponent term would be useful to 
stabilize the modulus during the inflation if 
we have a supergravity inflation model with the 
inflation $Z$.
In the next section, we consider the inflation model, 
where the modulus is the inflaton.

\section{Application example: racetrack inflation}

Here, we show one of application examples of 
the superpotential (\ref{eq:w-tot}), 
discussing the inflation model, where the modulus is 
the inflaton.
We consider the following total superpotential,
\begin{equation}
W_{\rm tot} = W_0 + \sum_{a=1,2} A_a e^{-\frac{2\pi}{N_a}(m_aS_0 + w_a T)}
+ A_3 e^{-\frac{2\pi}{N_3}{(m_3S_0 - w_3 T)}},
\label{eq:W-total-7}
\end{equation}
with $m_{1,2} \geq 0,m_3, w_a >0$ for $a = 1,2,3$.
When $A_3 =0$, the above superpotential (\ref{eq:W-total-7}) 
corresponds to one in the racetrack inflation 
model \cite{BlancoPillado:2004ns}.
(See also Ref.~\cite{Lalak:2005hr}.)
Actually, when we choose parameters appropriately, e.g. 
\cite{BlancoPillado:2004ns}
\begin{eqnarray}
 & & \frac{N_1}{w_1} = 100, \qquad \frac{N_2}{w_2} = 90, \qquad 
W_0 = -\frac{1}{25000}, 
\nonumber \\
 & & A_1e^{-\frac{2\pi}{N_1}m_1S_0 }= 
\frac{1}{50}, \qquad A_2e^{-\frac{2\pi}{N_2}m_2S_0} = -
\frac{35}{1000}, 
\end{eqnarray}
there is a saddle point at $(Re(T),Im(T))=(123.22,0)$ and 
the two minima $(Re(T),Im(T))=(96.130,\pm 22.146)$ for $A_3=0$.
Around the saddle point, the slow-roll inflation can be realized.
That is, at the saddle point, we obtain $\varepsilon =0$ and 
$\eta = -0.006097$.
The slow-roll parameters $\varepsilon$ and $\eta$ are 
defined as 
\begin{equation}
\varepsilon \equiv \frac{M_p^2}{2}\frac{1}{V^2}
\left( \frac{dV}{d\phi} \right)^2, \qquad 
\eta \equiv M_p^2 \frac{1}{V}\frac{d^2V}{d^2\phi},
\end{equation}
for the canonically normalized inflaton $\phi$.


\begin{figure}[t]
\begin{center}
\begin{minipage}{0.48\linewidth}
\epsfig{figure=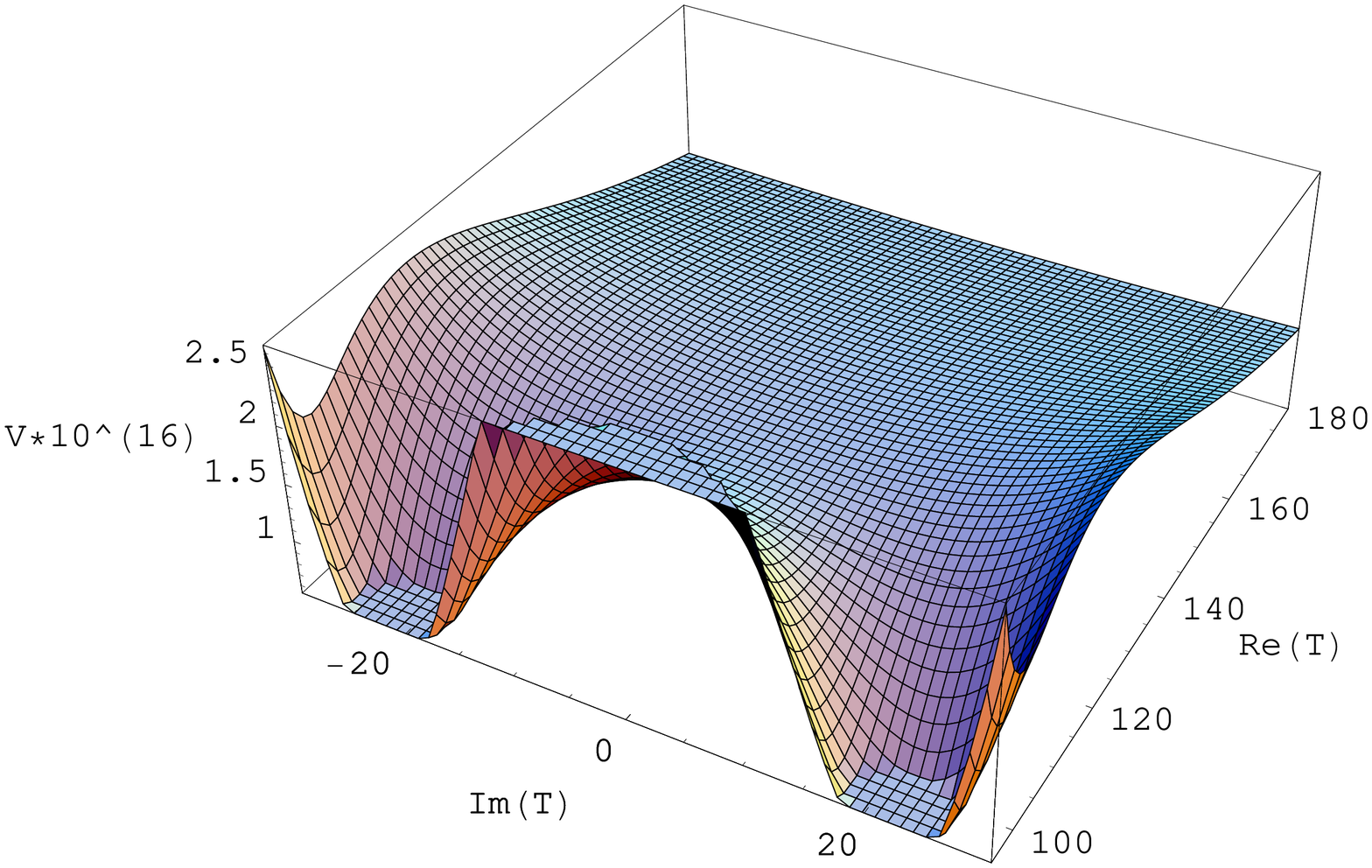,width=\linewidth}
\centerline{(a)}
\end{minipage}
\hfill
\begin{minipage}{0.48\linewidth}
\epsfig{figure=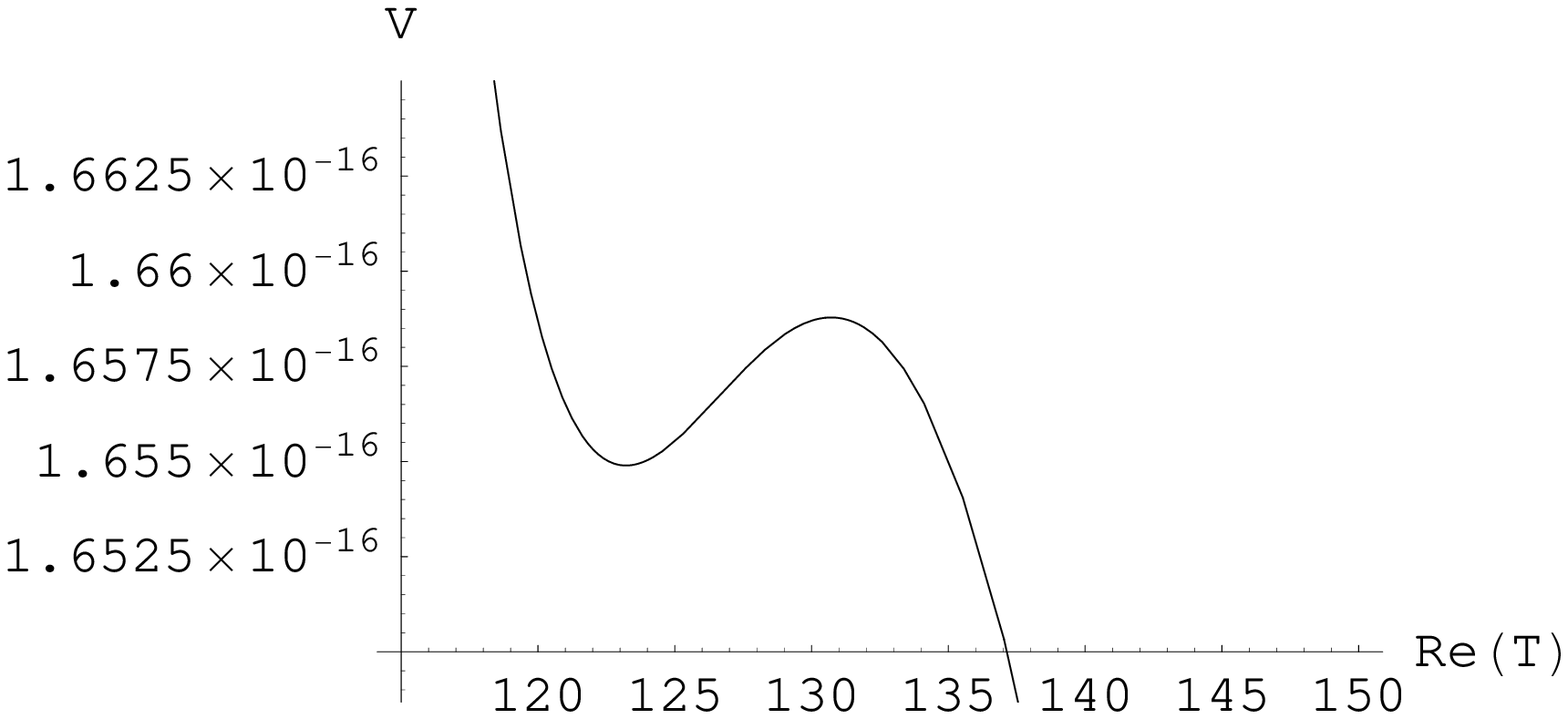,width=\linewidth}
\centerline{(b)}
\end{minipage}
\end{center}
\caption{The graph (a) shows a racetrack type scalar potential derived from 
Eq.~(\ref{eq:W-total-7}) with $A_3 =0$
(rescaled by $10^{16}$). Inflation begins in a vicinity of the saddle point
at $Re(T) = 123.22,~Im(T)=0$. The graph (b) shows a slice of the potential along $Im(T)=0$. }
\label{racetrack}
\end{figure}

Fig.\ref{racetrack} (a) shows the scalar potential around the saddle point for $A_3 =0$, and the section
along the direction $Im(T) =0$ is shown in Fig.\ref{racetrack} (b).
The bump around $Re(T) \approx 130$ is low and 
the modulus $Re(T)$ may overshoot the saddle point and 
run away to infinity.


\begin{figure}[t]
\begin{center}
\begin{minipage}{0.48\linewidth}
\epsfig{figure=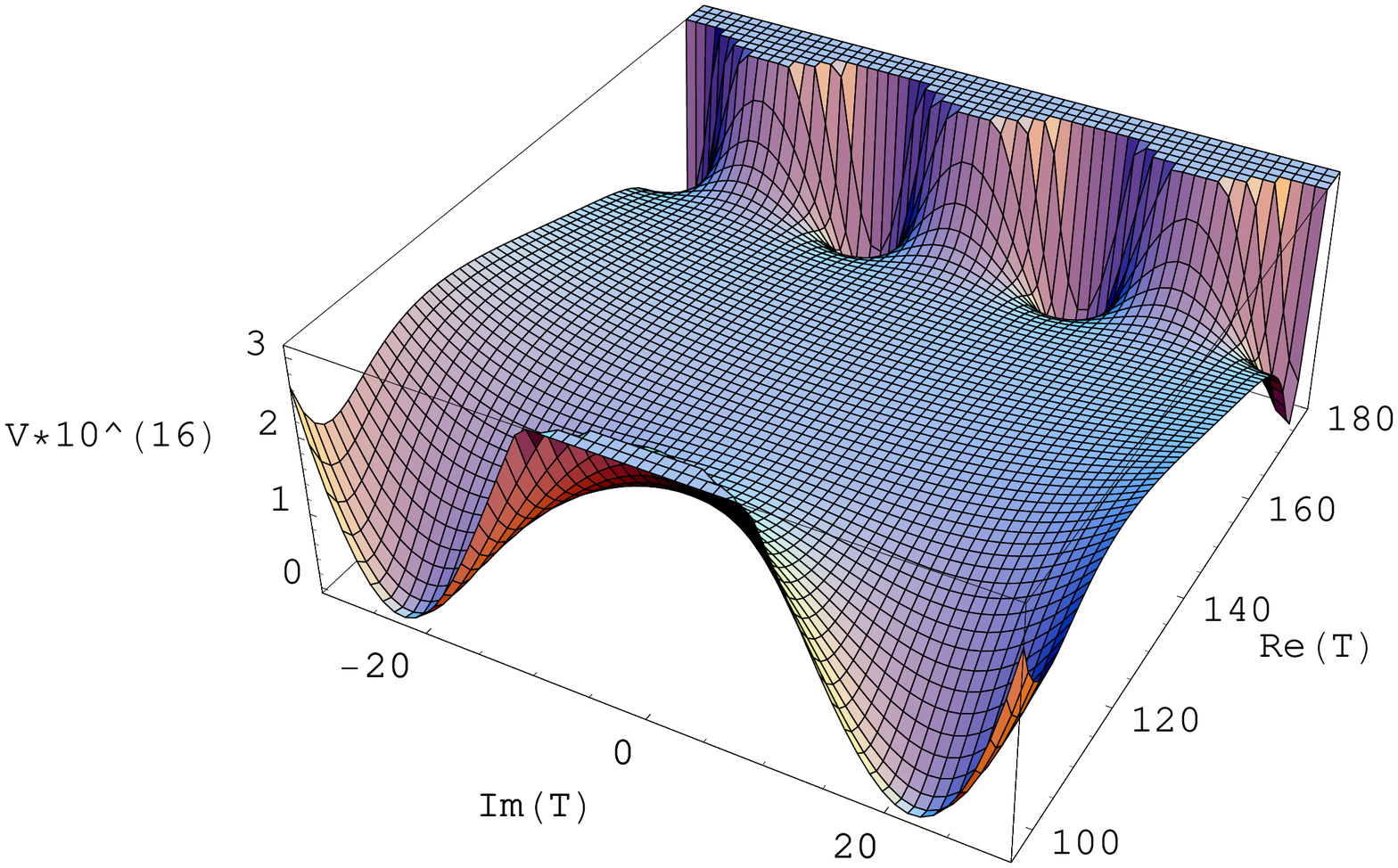,width=\linewidth}
\centerline{(a)}
\end{minipage}
\hfill
\begin{minipage}{0.48\linewidth}
\epsfig{figure=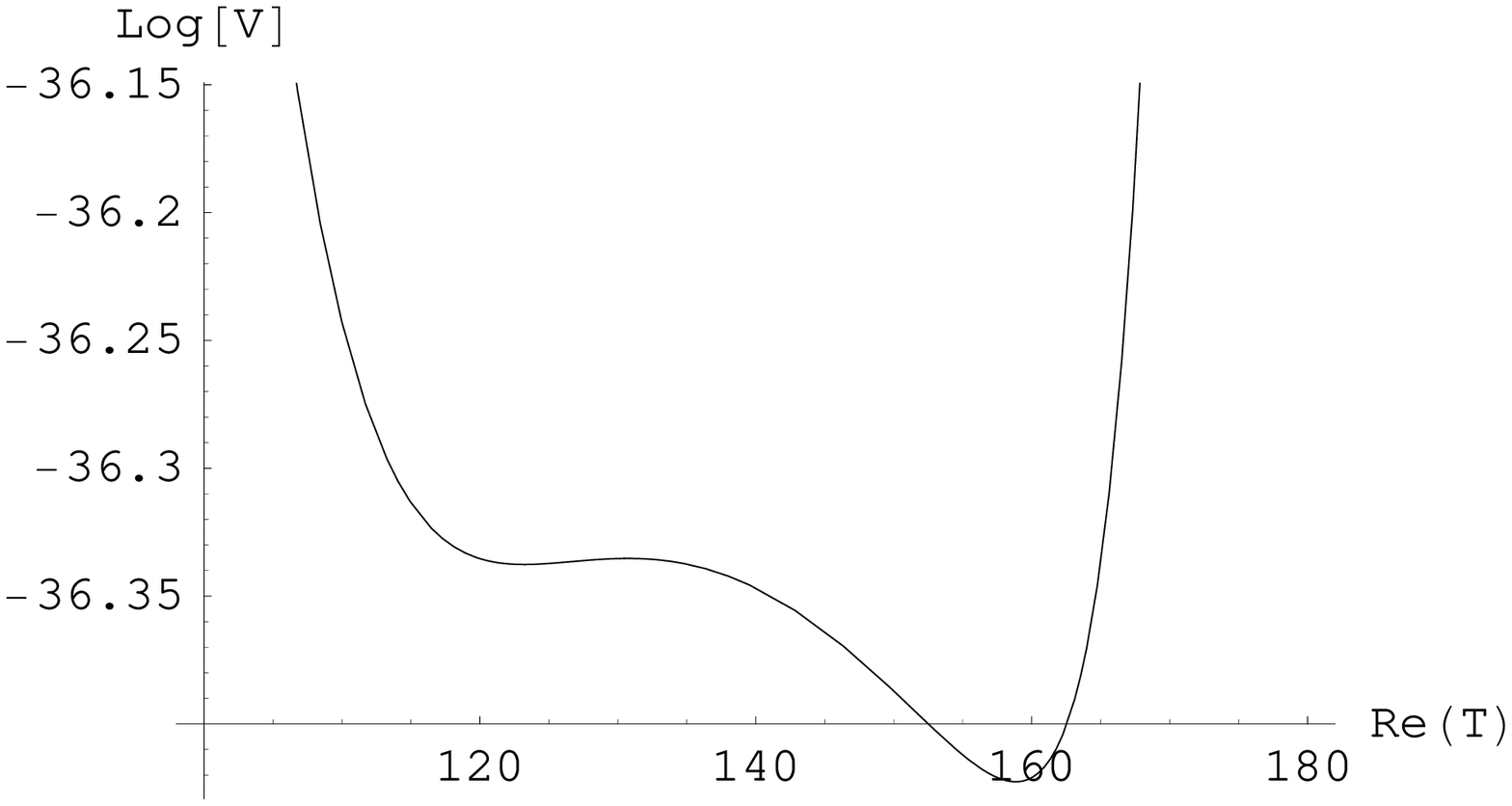,width=\linewidth}
\centerline{(b)}
\end{minipage}
\end{center}
\caption{The graph (a) shows a racetrack type scalar potential 
derived from Eq.~(\ref{eq:W-total-7}) with $A_3 =1 $
(rescaled by $10^{16}$). Inflation begins in a vicinity of the saddle point
at $Re(T) = 123.22,~Im(T)=0$ and we can find new minima which are shallower than old ones around $Re(T) \approx 173,~Im(T) \approx 10$
and high barrier around $Re(T) \approx 210$. 
The graph (b) shows a slice of the natural logarithm of the potential along $Im(T)=0$.}
\label{racetrack_posi}
\end{figure}

Now, we add the third term of the superpotential 
(\ref{eq:W-total-7}) such that it lifts up the runaway direction 
without violating the potential behavior around the saddle point.
For example, when we tune the parameters as\footnote{Then this potential is valid for  $ 68.8\pi /m_3 < Re(T) < 68.8\pi $.} 
\begin{equation}
A_3 =1, \qquad
m_3S_0 = 68.8\pi, \qquad w_3 =1, \qquad  N_3= 20,
\end{equation}
the scalar potential becomes as shown in Fig.~\ref{racetrack_posi}.
The height of the barrier around $Re(T) \approx 210$ is 
of ${\cal O}(M_p^4)$ and that would be helpful to 
avoid the overshooting problem.
In addition, at the saddle point, we obtain 
$\varepsilon =0$ and $\eta = -0.006850$, 
that is, the potential behavior around the saddle point, 
which is important to realize the inflation, does not change.
Indeed, before the inflaton rolls down to the minimum 
$(Re(T),Im(T))=(96.130,22.146)$, we can realize $N=130$ e-foldings 
when we use the initial conditions $(Re(T),Im(T))=(123.22,0.1)$, 
which is a vicinity of the saddle point.
This number of e-folds is almost the same as one obtained in 
Ref.~\cite{BlancoPillado:2004ns}.
Thus, adding the superpotential term with positive exponent 
is useful to lift up the runaway direction.

Furthermore, we could construct other racetrack inflation models 
with the following superpotential,
\begin{equation}
W = W_0 + Ae^{\pm aT} + B e^{bT},
\end{equation}
with $a, b >0$.
We would study such possibility elsewhere \cite{newwork}.

\section{Conclusion and discussion}

We have studied moduli superpotentials with 
positive exponents.
The corresponding scalar potentials have 
a quite rich structure.
There are several local minima with different 
potential energies and a high barrier of ${\cal O}(M_p^4)$ 
as well as the runaway vacuum.
This form of the scalar potentials has
significant implications from the viewpoints of 
cosmology and particle phenomenology, e.g. 
the realization of inflation models, avoiding 
the overshooting problem and destabilization due to 
finite temperature effects.
This type of potentials would be useful to 
realize the inflation and low-energy supersymmetry breaking.
Thus, it would be interesting to study a new type of 
inflation models with positive exponent terms.
In addition, we have shown that the modulus can be 
stabilized by a single term superpotential with 
positive exponent.
That is one of the simplest models for the 
modulus stabilization and SUSY breaking.
It would be interesting to apply this form of the potential 
to several phenomenological and/or cosmological aspects, 
e.g. the QCD axion.

\subsection*{Acknowledgement}
H.~A.\/,  T.~H.\/ and T.~K.\/ are supported in part by the
Grand-in-Aid for Scientific Research No.~182496, No.~194494 
and  No.~20540266 from the Ministry of Education, Culture,
Sports, Science and Technology of Japan.
O.~S.\/ is supported by the MEC project FPA 2004-02015 and 
the Comunidad de Madrid project HEPHACOS (No.~P-ESP-00346).

\end{document}